# Efficient infrared upconversion via a ladder-type atomic configuration

Dong-Sheng Ding, Zhi-Yuan Zhou, Bao-Sen Shi[*], Xu-Bo Zou

*Key Laboratory of Quantum Information, University of Science and Technology of China, Hefei 230026, China*

We experimentally demonstrate infrared light at 1529.4nm can be converted into the visible at 780nm with 54% efficiency through a ladder-type atomic configuration in $^{85}$Rb. In particular, theoretically, we analyze the high efficiency is due to the large nonlinear dispersion of the index of refraction from the off-resonant enhancement in four-wave mixing process. By using two perpendicular polarized pump fields, the coherence of two FWM processes in this configuration is verified.

Four-wave mixing (FWM) is a nonlinear optical process in which three fields are coherently interacted to generate a fourth field in a medium. This nonlinear process has an important application: efficient field up-conversion from the infrared spectrum to the visible [1]. The nonlinear parametric interaction processes in atoms with substantial efficiency should satisfy the phase matching condition. Efficient frequency up-conversion should occur within phase matching lengths, otherwise, the generated field decrease rapidly with the polarization produced by the signal fields, resulting in destructive interference and low conversion [2].

FWM with high efficiency in a double $\Lambda$ system [3, 4, 5] or a diamond configuration [6] has been obtained. Recently, Hsu *et al.* have demonstrated the generation of a field with a conversion efficiency of 5% in a double-ladder system through resonant FWM in a hot atomic vapor cell [7], and this process is theoretical analyzed in detail in Ref. 8 and 9. In these works, two lasers with wavelength of 794.3 nm and a laser with wavelength of 852.2 nm are used to generate a field at 852.2 nm. Willis et al. performed a FWM experiment in a diamond-type atomic level configuration in a warm rubidium (Rb) vapor [10]. In their experiment, three lasers with wavelengths of 780 nm, 795 nm and 1324 nm were used to generate a field at 1367 nm. The maximum conversion efficiency they achieved was about $1.3 \times 10^{-4}$. So the high efficiency of 54% achieved in hot atomic vapor cell using a ladder-type configuration in our experiment is an important advance in nonlinear optical process. Our experiment is performed using a Ladder-type in hot atoms, which is different to Ref. 6 where the complex cooling system is used. In addition, we use this ladder-type configuration to perform correlated photon pairs generated experiment [11], cascade image transfer [12] and linear conversion of orbital angular momentum [13].

A problem needed to be resolved is how to enhance the nonlinear susceptibility and eliminate resonant absorption of the atom to the generated field in FWM. Ref. [4] reports a work of enhancing nonlinear susceptibility and reducing linear susceptibility using coherent properties. In this report, we demonstrate an up-conversion nonlinear wave mixing process with high efficiency in hot $^{85}$Rb atomic vapor cell, where the infrared field at 1529.4nm is converted to the visible field at 780nm. Within our configuration, the frequencies of all fields are off-resonant with respect to the atomic transitions. Hence, the generated field decouples to atoms and can be radiated from atoms freely. In addition, the high efficiency we obtained is function of two-photon detuning, and the optimal efficiency appears near to two-photon resonance according to experimental result. The two-photon near-resonance helps the upper level and the ground level form a coherent population trapping state (CPT) through only two pump fields. But this state would be decreased due to the presence of signal field, which make the population of upper level and ground level redistributed. Thereby, even if the nonlinear process appears under the conditions when complete CPT does not occur, but the nonlinear susceptibility in this process can be enhanced. We give a theoretical result which is in reasonable agreement with our experimental data.

The ladder-type configuration of $^{85}$Rb is shown by Fig.1 (a). It consists of one ground state |3>, one intermediate states |2> and one upper state |1>. In our experiment, the ground state is $5S_{1/2}(F=3)$, the intermediate states are $5P_{3/2}$ and the upper state is $4D_{5/2}$. The transition frequency between the ground state and the intermediate states is the D2 line (780 nm) of $^{85}$Rb, and the transition between the intermediate state and the upper state can be coupled with a laser at 1529.4 nm. We experimentally generate a field at 780 nm using the pump 1, pump 2 and signal via a nonlinear FWM process.

The simplified experimental setup is shown in Fig. 1 (b). The cw laser at 780 nm (LD1, DL100, Toptica) was frequency-detuned $\Delta_1$ to the $^{85}$Rb atomic transition, $5S_{1/2}(F=3)$-$5P_{3/2}(F'=4)$. The remaining output from LD1 was used as the input pump 2 to a 5-cm-long cell filled with $^{85}$Rb. A cw laser at 1529 nm (LD2, DL 100, Prodesign, Toptica) was divided into two pump beams: pump 1 and signal, by using a beam splitter. These two beams both have vertical polarization directions and nearly co-propagate through the Rb cell. The pump 1 nearly counter-propagates with the pump 2 beam by the angle $\beta = 0.37°$, the pump 2 has circular polarization. The upconverted field at 780 nm with the same polarization with the pump 2 nearly counter-propagates with the signal by the angle $a = 0.61°$, and is monitored by a power meter. The signal and the pump 2 are focused and the diameters at the center of the Rb cell are approximately 185 $\mu m$ and 195 $\mu m$ respectively. The pump 1 is weakly focused and the beam waist is about 1 mm in the cell. LD2 was frequency-detuned $\Delta_2$ to the $^{85}$Rb atomic transition, $5P_{3/2}(F'=4)$-$4D_{5/2}(F''=4)$.

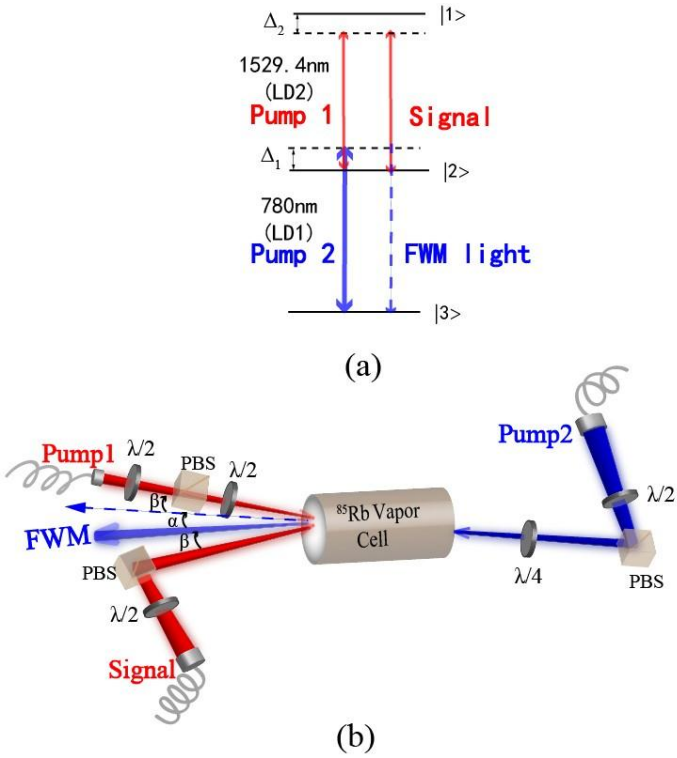

FIG. 1 (Color online)(a) Energy levels diagram of ladder-type configuration (b) The experimental setup.

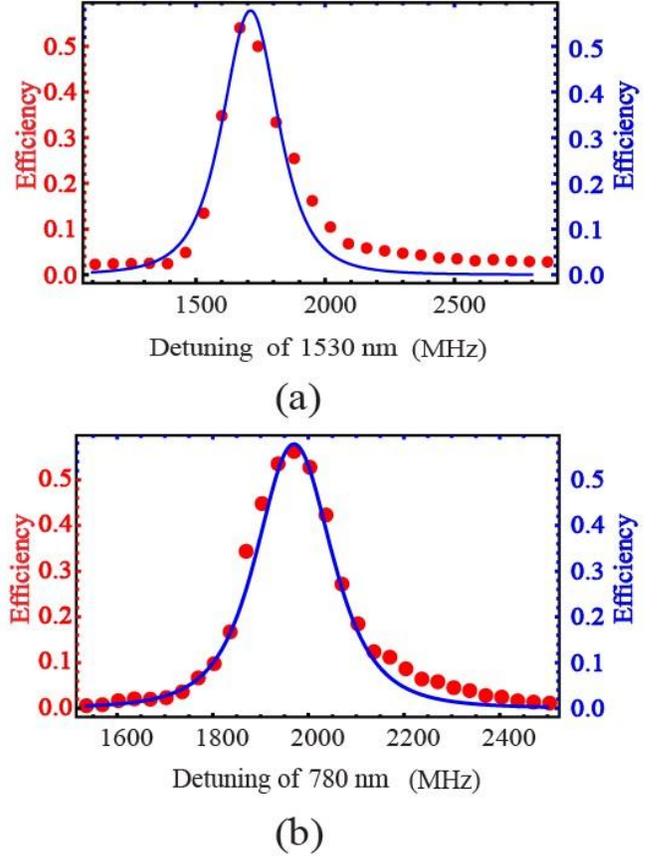

FIG. 2. (Color online) (a) and (b) Conversion efficiencies against the detuning of the fields at 780 nm and 1530nm. The red dots are experimental data, blue line is calculated results. Relevant parameters for calculations are $\Delta_1 = 2.0$ GHz, $\Delta_2 = 1.67$ GHz, $\Omega_{p2} = 590$ MHz, $\Omega_{p1} = 98$ MHz, $\Omega_s = 31$ MHz. $\gamma = 6$ MHz. The calculated curves consider Doppler broadening at $200\,^\circ C$ and the light shift.

We use a power meter to measure the power of the generated FWM signal. The powers of the pump 1, pump 2 and the signal are 9.6, 82 and 0.92 $mW$ corresponding to Rabi frequencies approximately are 98, 590 and 31 MHz respectively. The temperature of the cell is approximately $200\,^\circ C$. In the experiment, we find that the conversion efficiency is also strongly dependent on the single-photon detuning. The results are shown in Fig. 2. In Fig. 2a, the detuning $\Delta_1$ is fixed at about 2 GHz. In Fig. 2b, the detuning $\Delta_2$ is fixed at about 1.67 GHz. The detuning of $\Delta_1$ and $\Delta_2$ satisfied the two-photon near-resonance, where the atoms with the maximum probability velocity of $v = 304 m/s$ are populated and the light shift caused by large powers of pumps is considered. The maximal efficiency is about 54%, almost equal to the value based on a cold ensemble reported in Ref. 6. The main reason why the conversion efficiency is improved greatly compared with the reported works in Ref. 8 and 14, is mainly due to the use of the large single-photon detuning in our experiment. This can significantly reduce the resonant absorptions of the $^{85}$Rb atoms to the infrared light and the up-converted FWM signal. Our experimental results are in agreement well with the theoretical analysis given in Ref. 1, in which, large detuning of the pump 1 and the pump 2 is needed for high conversion efficiency. The noise is mainly from pump 2 field due to the pattern formation with high power [15].

It is worth illustrating that the polarization of the pump fields and signal field affect obviously the efficiency of conversion. We have done this experiment by changing the polarization of pump 2 with the vertical polarization pump 1, and change the polarization of pump 1 with the circular polarization pump 2. The optimal efficiency conversion process appears when the pump 2 has circular polarization and both the pump 1 and signal have vertical polarization shown by Fig. 3 ($\theta_{\lambda/2} = \theta_{\lambda/4} = 45^\circ$). This is due to the different transition routes or transition coefficients from different polarizations leading to different efficiency conversion. In this process, the frequencies of lasers are weakly unstable, but this can't hinder us to observe this main phenomenon along changing of the polarization.

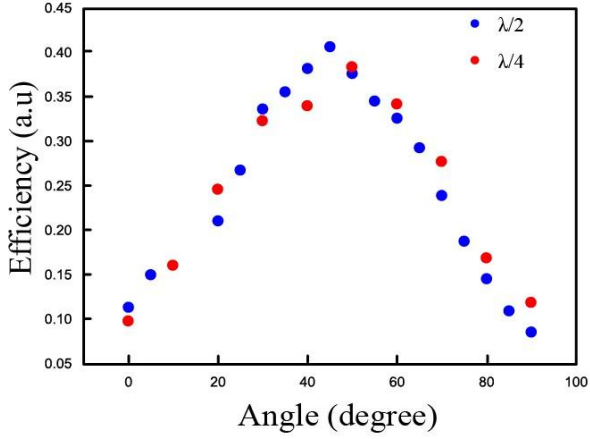

FIG. 3 (Color online) The efficiency against the degree of the $\lambda/2, \lambda/4$ plates inserted in pump 1 and pump 2 respectively. Blue dots are changing the polarization of pump 1 with the circular polarization pump 2, red dots are changing the polarization of pump 2 with the vertical polarization pump 1.

When the angle $\theta_{\lambda/4} = 45°$, the pump 2 can be treated as a vertical polarized and a horizontal polarized pump effects which induce two FWM processes. This FWM process is a coherent process, the generated vertical polarized FWM signal (VS) and horizontal polarized FWM signal (HS) are coherent. The results are shown by Fig. 4. Fig, 4(a) is the intensity of VS and HS. By using a polarizer, the coherence of VS and HS is obtained.

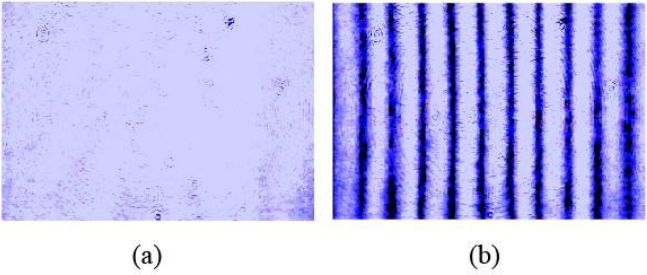

Fig. 4(Color online) (a) The simple intensity added with VS and HS. (b) The interference of VS and HS.

Next, we give a simply theoretical description of the FWM process in our system. The density matrix equation of motion referenced by [16] is

$$\frac{d\rho}{dt} = -\frac{i}{\hbar}[H, \rho] - \Gamma_{ij}\rho_{ij} \quad (1)$$

where, H is the effective interaction Hamiltonian; $\Gamma_{ij}$ ($i \neq j$) describes the complex decay rate from $|i\rangle$ to $|j\rangle$, $\Gamma_{ij}$ ($i = j$) describes the decay rate of $\rho_{ii}$. $\gamma_1, \gamma$ are the decay rates of the states $|2\rangle, |1\rangle$. We assume the Rabi frequency $\Omega_{p2} \gg \Omega_{p1}, \Omega_s, \Omega_{FWM}$, and approximately take $\gamma_1 = 6\gamma$ and drive the dynamical equation (1) in steady state. The FWM coupling equations are:

$$\begin{cases} \frac{\partial \Omega_s}{\partial z} = k_1 \Omega_s + k_2 \Omega_{FWM} \\ \frac{\partial \Omega_{FWM}}{\partial z} = k_4 \Omega_s + k_3 \Omega_{FWM} \end{cases} \quad (2)$$

Where, the coefficients $k_1, k_3$ are the linear terms corresponding to the linear susceptibility $\chi_s^{(1)}, \chi_{FWM}^{(1)}$, $k_2, k_4$ are nonlinear terms corresponding to the nonlinear susceptibility $\chi_s^{(3)}, \chi_{FWM}^{(3)}$. When considering the absorption of generated field, we obtain the simple formula $P_{FWM}^{(1)} = \frac{N\mu_{23}\Omega_{FWM}}{\varepsilon_0(I\gamma_1 + \Delta_1)}$, which tends to be small in the large detuning of $\Delta_1 = 2$ GHz. $\mu_{23}$ is the effective dipole moment with the transition $|3\rangle - |2\rangle$. The nonlinear susceptibility $P_{FWM}^{(3)}$ we calculated is:

$$P_{FWM}^{(3)} = \frac{N\mu_{23}}{\varepsilon_0} \times \frac{18i\gamma(4(\Delta_1 + \Delta_2)^2 - \Omega_{p2}^2)\Omega_{p2}\Omega_{p1}\Omega_s}{(12\gamma^2 + 3\Delta_1^2 - 4\Omega_{p2}^2)(9\gamma^2(\Delta_1 + \Delta_2)^2 + (-(\Delta_2(\Delta_1 + \Delta_2)) + \Omega_{p2}^2)^2)}$$

(3)

In our system, the nonlinear susceptibility $P_{FWM}^{(3)}$ is $10^4$ times larger than the linear susceptibility $P_{FWM}^{(1)}$ at large detuning $\Delta_1, \Delta_2$. So, we can simplify Eq. (2) by ignoring the linear terms. The FWM field can be easily represented as $\vec{E}_{FWM} = \chi^{(3)} \vec{E}_{p1} \vec{E}_{p2} \vec{E}_s$. By calculating the third-order susceptibility, we obtain the efficiency $\eta = E_{FWM}^2 / E_s^2$ and fit our experimental results shown by Fig. 2 (a) and Fig. 2 (b). We obtain that the results are reasonable in agreement with our experiment data. In Fig. 2 (a) and Fig. 2 (b), the efficiency at the edge of experimental data is not zero due to the noise from pump 2 which is hard to reduce. We make our best efforts to reduce this drawback which is reason why this high efficiency conversion process can't be obtained at photon level in our configuration. Ref. 6 gives a high efficiency nonlinear process at high power level or photon level where the filters are used to obstruct noise. The interference intensity is obtained : $|\vec{E}_{VS+HS}|^2 = |\chi^{(3)} \vec{E}_{p1} \vec{E}_s (\vec{E}_{p2V} + \vec{E}_{p2H})|^2$. The intensity is interferences fringes due to the coherence of vertical polarized and horizontal polarized pump 2 fields.

In conclusion, we demonstrate a considerable frequency conversion via FWM process based on the atomic ladder-type configuration. We find this nonlinear process affected by the detuning significantly, and give a theoretical prediction which is agreement with our experimental data. The large detuning is the key of high efficiency conversion in our system, and the polarization also affects frequency up-conversion. In addition, we verify that the two FWM processes with two perpendicular polarized pump fields are coherent.

**Acknowledgements**

This work was supported by the National Natural Science Foundation of China (Grant Nos. 61275115, 11174271), the National Fundamental Research Program of China (Grant No. 2011CB00200), the Innovation fund from CAS, Program for NCET.

[*] *drshi@ustc.edu.cn*